\documentclass[acmlarge,nonacm]{acmart}

\usepackage[ruled]{algorithm2e}
\usepackage{graphicx}
\usepackage{subfigure}
\usepackage{url}
\usepackage{booktabs}
\usepackage{multirow}
\usepackage{subfigure}

\begin{document}

\title{Estimating Sunlight Using GNSS Signal Strength from Smartphone}


\author{Yuuki Nishiyama}
\orcid{0000-0002-5549-5595}
\email{yuukin@iis.u-tokyo.ac.jp}
\affiliation{%
 \institution{The University of Tokyo}
 \city{Meguro}
 \state{Tokyo}
 \country{Japan}}
 
\author{Kosuke Hatai}
\email{hatai@mcl.iis.u-tokyo.ac.jp}
\affiliation{%
 \institution{The University of Tokyo}
 \city{Meguro}
 \state{Tokyo}
 \country{Japan}
 \postcode{153-8505}
}

\author{Kota Tsubouchi}
\email{dongxuefu@mcl.iis.u-tokyo.ac.jp}
\affiliation{%
 \institution{Yahoo Japan Corporation}
 \city{Chiyoda}
 \state{Tokyo}
 \country{Japan}}

\author{Kaoru Sezaki}
\orcid{0000-0003-1194-4632}
\email{sezaki@iis.u-tokyo.ac.jp}
\affiliation{%
 \institution{The University of Tokyo}
 \city{Meguro}
 \state{Tokyo}
 \country{Japan}}

\renewcommand{\shortauthors}{Nishiyama, et al.}


\begin{abstract}
Excessive or inadequate exposure to ultraviolet light (UV) is harmful to health and causes osteoporosis, colon cancer, and skin cancer. The UV Index, a standard scale of UV light, tends to increase in sunny places and sharply decrease in the shade. A method for distinguishing shady and sunny places would help us to prevent and cure diseases caused by UV. However, the existing methods, such as carrying UV sensors, impose a load on the user, whereas city-level UV forecasts do not have enough granularity for monitoring an individual’s UV exposure. This paper proposes a method to detect sunny and shady places by using an off-the-shelf mobile device. The method detects these places by using a characteristic of the GNSS signal strength that is attenuated by objects around the device. As a dataset, we collected GNSS signal data, such as C/N0, satellite ID, satellite angle, and sun angle, together with reference data (i.e., sunny and shady place information every minute) for four days from five locations. Using the dataset, we created twelve classification models by using supervised machine learning methods and evaluated their performance by 4-fold cross-validation. In addition, we investigated the feature importance and the effect of combining features. The performance evaluation showed that our classification model could classify sunny and shady places with more than 97\% accuracy in the best case. Moreover, our investigation revealed that the value of C/N0 at a moment and its time series (i.e., C/N0 value before and after the moment) are more important features. 

\end{abstract}



\maketitle

\section{Introduction}
\label{sec:introduction}

Moderate natural sunlight is indispensable for all life on earth, while excessive or underexposure to sunlight can adversely affect the human body. Therefore, monitoring the duration of exposure to sunlight is useful for preventing diseases caused by over or under exposure to sunlight. Sunlight contains ultraviolet light (UV), and excessive exposure to UV is widely known to lead to serious diseases such as skin cancer, wrinkles, and cataracts~\cite{imokawa2015biological,narayanan2010ultraviolet}. On the other hand, in modern society, opportunities for getting direct sunlight are decreasing due to indoor lifestyles, underground mobility, and nightlife. Vitamin D is produced by the human body only when it is exposed to ultraviolet light; thus, underexposure to sunlight leads to vitamin D deficiency which increases the risk of a variety of diseases, including osteoporosis, colon cancer, and depression~\cite{grant2006epidemiology,nakamura2006vitamin,bis,bis2,hanley2005vitamin,scharla1998prevalence,yoshimura2013profiles,humble2010vitamin,focker2017vitamin,cannn}. 

Mobile devices such as smartphones have become widely popular in recent years and are carried by people on a daily basis. These mobile devices have rich sensing modules (e.g., location sensor, motion sensor, camera, and microphone). 
The analysis of sensor data passively collected from mobile devices is known to be able to detect various health-realted contexts~\cite{imwut_studentlife,imwut_drinking_episodes,imwut_wifi_respiration}.
Measuring the time spent in sunlight by using only objects that everyone owns like a smartphone would allow us to predict or prevent diseases above without imposing any extra burden on users. 
Weather forecasts provide UV values on a city scale and track their variability over time. However, since the UV value significantly changes depending on the location, it is difficult to measure the duration that an individual may be exposed to sunlight only with the weather forecast. In addition, a mobile UV meter enables one to record UV data over the course of a day, but it is unrealistic to expect that many people would buy a meter and carry it at all times. On the other hand, the signal strength from satellites that is used by the Global Navigation Satellite Systems (GNSS) is attenuated in a similar way as UV by screening objects such as buildings, roofs, and thick clouds. This characteristic has been used for detecting several contexts (i.e., detecting indoor-and-outdoor, volcanic ashes, and cumulonimbus)~\cite{EtsukoKatsuta2011gps, larson2013new,zhang2019real,koch2016soil}, but it has not been utilized for detecting UV-related contexts. 

To address this challenge, we propose a method that allows us to detect sunny and shady places using a GNSS module on off-the-shelf smartphones that most people carry with them every day. Using a characteristic that the GNSS signal strength is attenuated by screening objects, our method allows us to detect sunny and shady places by monitoring the angle of the sun and the angle and signal strength of each GNSS satellite. We created a classification model using four days worth of GNSS-related data collected at five (four outdoor and one indoor) places. A performance evaluation showed that the model can classify shady and sunny places on a minute by minute basis with more than 90\% accuracy.

The contributions of this paper are as follows.
\begin{itemize}
\item We devised a method for detecting sunny and shady places using an off-the-shelf smartphone,
\item We created a tool for collecting raw GNSS signals and UV-related values and used it to collect a dataset consisting of four days’ worth of GNSS and UV-related data (sunny and shady place labeled data) \item We evaluated the performance of our method and found that it can classify places with more than 90\% accuracy
\item We discuss future applications of the proposed method
\end{itemize}

This paper is organized as follows: Section~\ref{sec:related_work} shows related work on methods for estimating UV exposure and context recognition using GNSS signals. The motivation of this paper and the basic idea for detecting shady and sunny places using GNSS signals are described in Section~\ref{sec:motivation}. Section~\ref{sec:approach} explains the implementation of our method and the dataset that we collected. The performance evaluation is in Section~\ref{sec:evaluation}. Section~\ref{sec:discussion} discusses the results of the evaluation and future applications. Section~\ref{sec:conclusion} concludes the paper.

\section{Related Works}
\label{sec:related_work}

Nowadays, many consumer devices (i.e., smartphones, smartwatch, and IoT devices) come equipped with a wireless communication module, and a number of context recognition methods have been proposed that use its signal characteristics. 
Since UV exposure has a significant impact on the human body, various measurement and estimation methods have been proposed. 

Section~\ref{sec:2_gnss} summarizes the methods of detecting contexts using signals strength, while section \ref{sec:2_uv} describes the UV detection and estimation methods.

\subsection{Context Recognition Using Signal Strength}
\label{sec:2_gnss}
The signal strength and its characteristics received by the wireless communication module (i.e., WiFi, Mobile network, and GNSS) have been used in various context recognition methods~\cite{imwut_maekawa_wifi, imwut_wifi_rssi,imwut_cellular_signal}.
For example, WiFi signal strength~\cite{imwut_maekawa_wifi,imwut_wifi_rssi} and cellular signaling data~\cite{imwut_cellular_signal} are utilized for indoor and outdoor localization, respectively.
Moreover, WiPhone~\cite{imwut_wifi_respiration} used WiFi Channel State Information (CSI) to estimate human breathing events. 

The status of GNSS signal reception has been used as a data source of context recognition in several reseraches~\cite{EtsukoKatsuta2011gps,larson2013new,zhang2019real,koch2016soil}. In particular, there have been several studies on determining whether the GNSS signal is blocked by obstacles on the basis of the received signal.
Katsuta et al. \cite{EtsukoKatsuta2011gps} took advantage of the fact that the signal-to-noise ratio (SNR) value differs greatly between indoor and outdoor locations, and by collecting SNR values in advance and building a model, they were able to automatically determine whether the user was outdoors or indoors. Their performance evaluation shows that their model can distinguish between indoors and outdoors with more than 90\% accuracy with only seven seconds’ worth of SNR values. Koch et al. \cite{koch2016soil} developed a method to measure soil moisture from GNSS sensor measurements. In addition, Larson et al. \cite{larson2013new} detected volcanic ejecta such as volcanic ash based on C/N0 values instead of the conventional method using radar or satellites. Zhang et al. \cite{zhang2019real} established a real-time method for estimating carrier multipath based on changes in C/N0 values.

\subsection{Monitoring and Estimating UV Exposure}
\label{sec:2_uv}
Detailed natural environmental data, such as temperature, air quality, and UV Index are important information to improve people's quality of life. However, it is impossible to granularly measure the data from all over the world using specific measurement tools in real-time.

To address the challenge, methods are proposed to estimate environmental conditions in different granularity such as global~\cite{uv_global,arola2002assessment}, city~\cite{imwut_air_quality}, and personal~\cite {schmalwieser2020possibilities} scale. 
Especially, estimation of environmental conditions on a personal scale is important information in managing a person's physical and mental condition, and it helps prevent various diseases. In this paper, from the environmental data, we focus on UV exposure that makes the various reason of various diseases such as skin cancer, wrinkles, cataracts, colon cancer, and depression ~\cite{imokawa2015biological,narayanan2010ultraviolet,grant2006epidemiology,nakamura2006vitamin,bis,bis2,hanley2005vitamin,scharla1998prevalence,yoshimura2013profiles,humble2010vitamin,focker2017vitamin,cannn}.

Miyauchi et al. \cite{miyauchi2016determining} conducted a detailed simulation to estimate the amount of UV exposure sufficient to maintain health. They used the SMART2 model to simulate ozone and other aspects of the atmosphere to estimate the amount of erythemal UV radiation that causes sunburn and the amount of UV radiation needed to produce vitamin D. Both UV radiation data and the UV Index were used to estimate the amount of UV radiation needed to produce vitamin D. They concluded that both types of UV radiation were almost linearly correlated with the UV Index and that the UV Index could be used to determine the duration of exposure to UV radiation necessary to maintain health. 
%
Schmalwieser et al. \cite{schmalwieser2020possibilities} developed a tool, called EFTA, to estimate the UV dose associated with an individual based on measurements of the ambient UV dose. It is calculated by multiplying by a factor depending on people's behavior. However, this method cannot be used in general yet because it requires a measurement of the ambient UV radiation and can only determine the UV exposure in the case of a certain height of the sun and a certain part of the body. 

Although various researches tackle estimating UV exposure,  a method to passively obtain the daily UV exposure using a device that people carry every day has not been proposed. 
\section{Basic Idea for Detecting UV-related Context Using GNSS Signal}
\label{sec:motivation}
The ultimate goal of this study is to estimate the amount of UV radiation exposure daily for maintenance of health. From Section~\ref{sec:related_work}, it can be seen that most of the studies on estimating UV exposure were based on ambient UV exposure data. This is accurate enough for a rough estimation of UV exposure. However, since the amount of UV exposure is greatly affected by the surroundings, it is difficult to estimate it in detail only from the surrounding data without the data on shielding in the environment. In other words, data on the surrounding shields are essential for estimating UV radiation exposure.

\begin{figure}[tbp]
    \centerline{\includegraphics[scale=0.4]{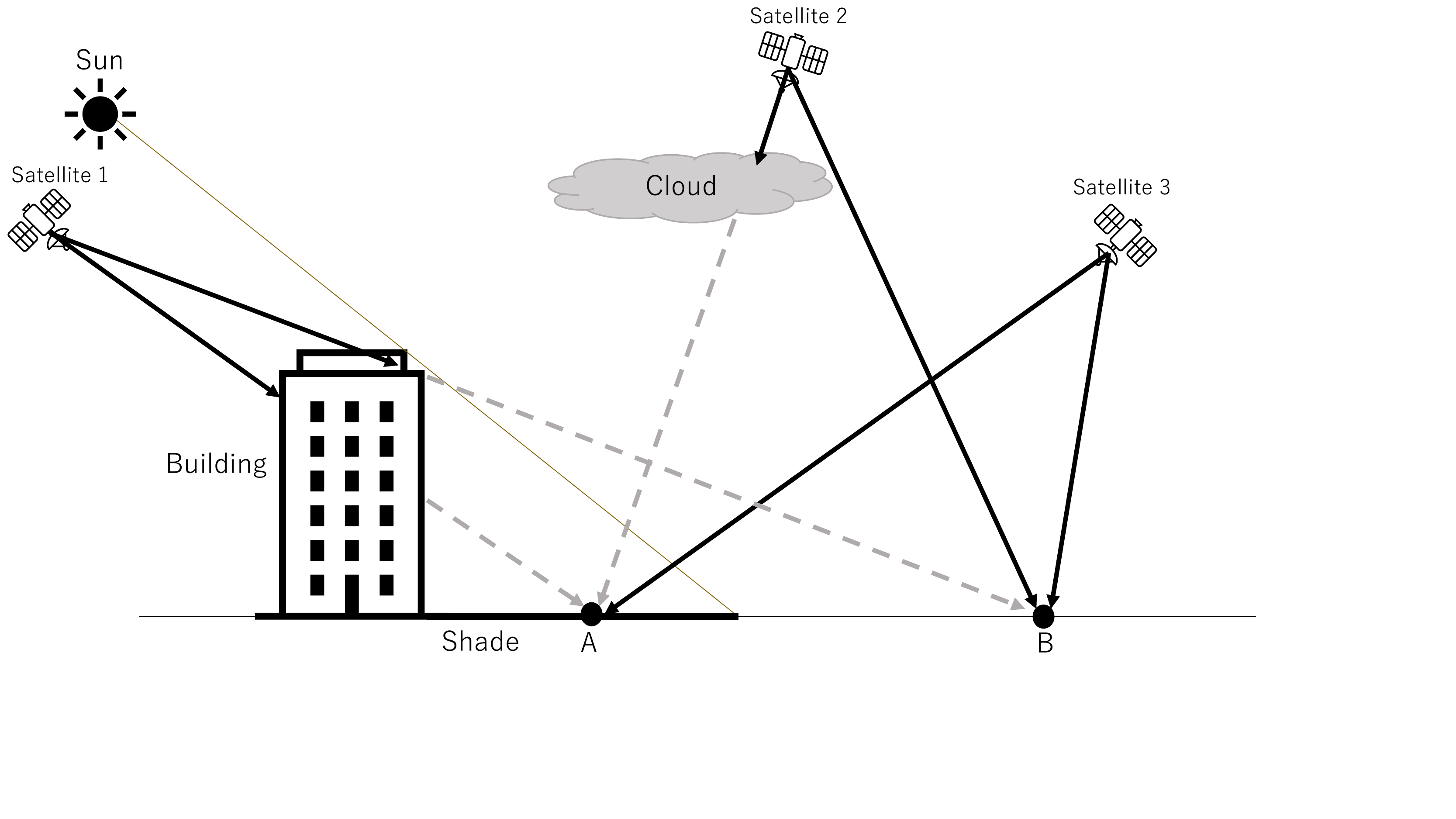}}
    \caption{Sunlight and GNSS signal strength}
    \label{fig:figure0}
\end{figure}

As described in Section~\ref{sec:related_work}, there are previous studies on detecting shields from the reception strength of GNSS signals, so we decided to use them as a basis.
To estimate the change in UV exposure due to shielding, we will focus on the distinction between sun and shade (fig.~\ref{fig:figure0}) and verify whether GNSS signal strength can be used in future methods of UV exposure estimation.

\subsection{Characteristics of UV Light}
UV rays are divided into three types according to their wavelength: UV-A (315 nm~400 nm), UV-B (280 nm~315 nm), and UV-C (100~280 nm)~\cite{UV2020}. In UV rays, shorter wavelengths are more toxic than longer wavelengths; thus, UV-C and UV-B are more toxic than UV-A. However, most of the UV-C and 90\% of the UV-B rays are absorbed by the ozone layer as they enter the atmosphere. Moreover, excessive UV exposure is known to cause various health problems such as skin cancer, blemishes, and wrinkles. On the other hand, exposure to UV-B is necessary for the production of vitamin D, which affects the occurrence of depression and osteoporosis.
The UV index is an index of the intensity of ultraviolet rays to indicate the degree to which they affect the human body in an easy-to-understand manner. The index is obtained by integrating intensity over wavelengths from 250 to 400 nm and then multiplying the result by 40 if the unit is $W/m^2$~\cite{UVI}. 
The WHO evaluates the UV intensity according to the UV Index value on five levels~\cite{world1995global}.


\subsection{Raw GNSS Data}
\label{sec:raw_gnss_data}
Global Navigation Satellite System (GNSS) is a generic term for satellite positioning systems, such as Global Positioning System (GPS) in the U.S., Quasi-Zenith Satellites (QZSS) in Japan, GLONASS in Russia, and Galileo in the European Union. GNSS enables us to estimate one's location using received signals from more than four satellites at the same time. The signal from the satellites contain information on the orbits of the satellites and the precise times of atomic clocks.

\begin{figure}[tbh]
    \centering
    \includegraphics[width=0.8\linewidth]{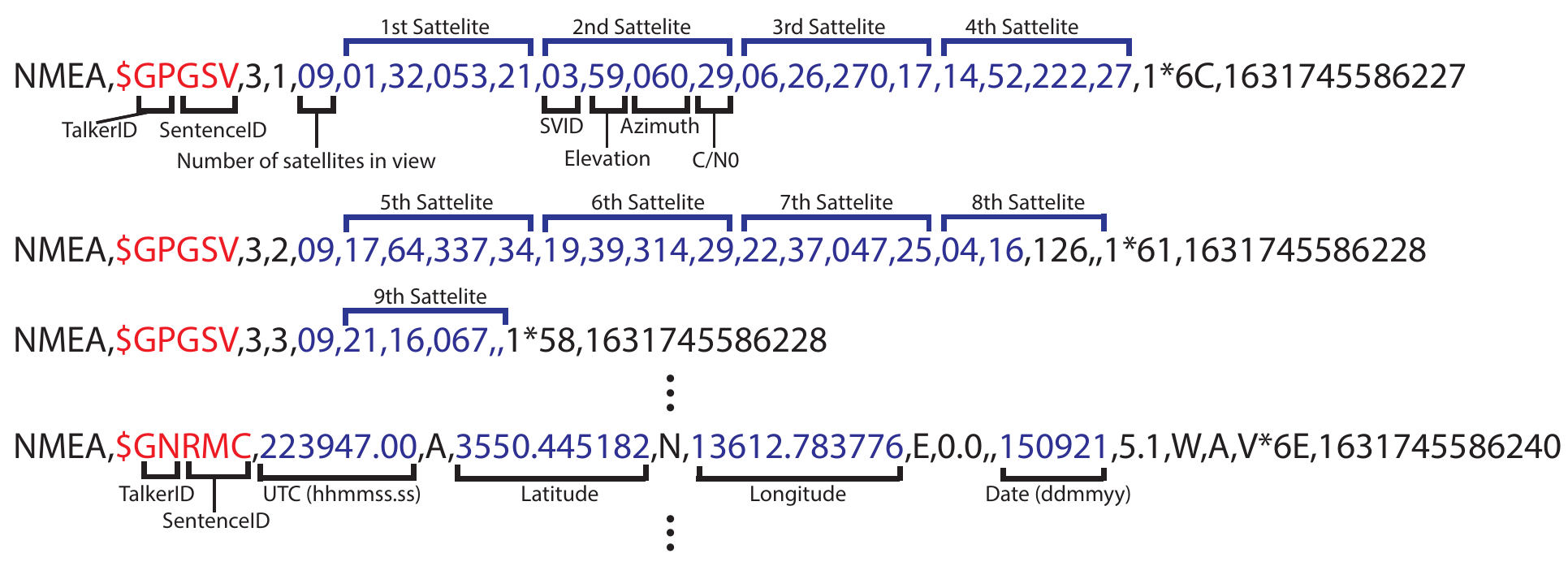}
    \caption{Raw GNSS data}
    \label{fig:figure451}
\end{figure}

Fig.~\ref{fig:figure451} depicts the raw data values obtained from the sensors. The raw GNSS data were saved in NMEA format, as shown in Fig.~\ref{fig:figure451}, and they contained a great deal of information. In this format, a line is a data record and each data is split by a comma. \verb|$GPGSV| means \verb|GNSS Satellites in View with satellites of GPS, SBAS, and QZSS|. The row contains the satellite ID (SVID), angle (Azimuth and Elevation), and signal strength (C/N0). Moreover, \verb|$GNRMC| means \verb| Recommended minimum specific GPS data calculated by satellites of GPS, QZS, and Galileo|. \verb|$GNRMC| contains the date, time, and location (latitude and longitude). Feature extraction was performed on six pieces of information: four from \verb|$GPGSV| (SVID, angle of azimuth, angle of elevation, and signal strength (C/N0)), and two from \verb|$GNRMC| (date and time, location).

Table~\ref{tab:raw_gnss_data} shows representative raw GNSS data. C/N0 indicates the GNSS signal reception strength, SVID indicates the satellite ID transmitting the GNSS signal, and satellite elevation and azimuth indicate the angle of the signal from the satellite.

\begin{table}[htb]
    \centering
    \caption{Representative Raw GNSS Data}
    \begin{tabular}{ l r r } 
    \toprule
        Data name & unit & format \\ 
    \midrule
        SVID (Sattelite ID) & - & int \\ 
        Time related data & - & string \\ 
        Latitude  & degree & float \\ 
        Longitude & degree & float \\ 
        Satellite elevation & degree & int  \\ 
        Satellite azimuth & degree & int  \\ 
        C/N0 & dBHz & int \\ 
    \bottomrule
    \end{tabular}
    \label{tab:raw_gnss_data}
\end{table}



\subsection{Definition of Sunny and Shady Places}
\label{subsec:def_sun}
In this study, we want to find a way of distinguishing between sunny and shady place. Outside, even in the shade, there may be sunlight and the UV Index would not be zero. Therefore, we need to make a clear distinction between sunny and shady places. We performed a preliminary experiment to determine UV-Index thresholds for sunny and shady places. We experimented with a large wall to examine how the values change when we moved from the shade into the sun and from in the sun into the shade.

\begin{figure}
    \centering
    \includegraphics[width=0.8\linewidth]{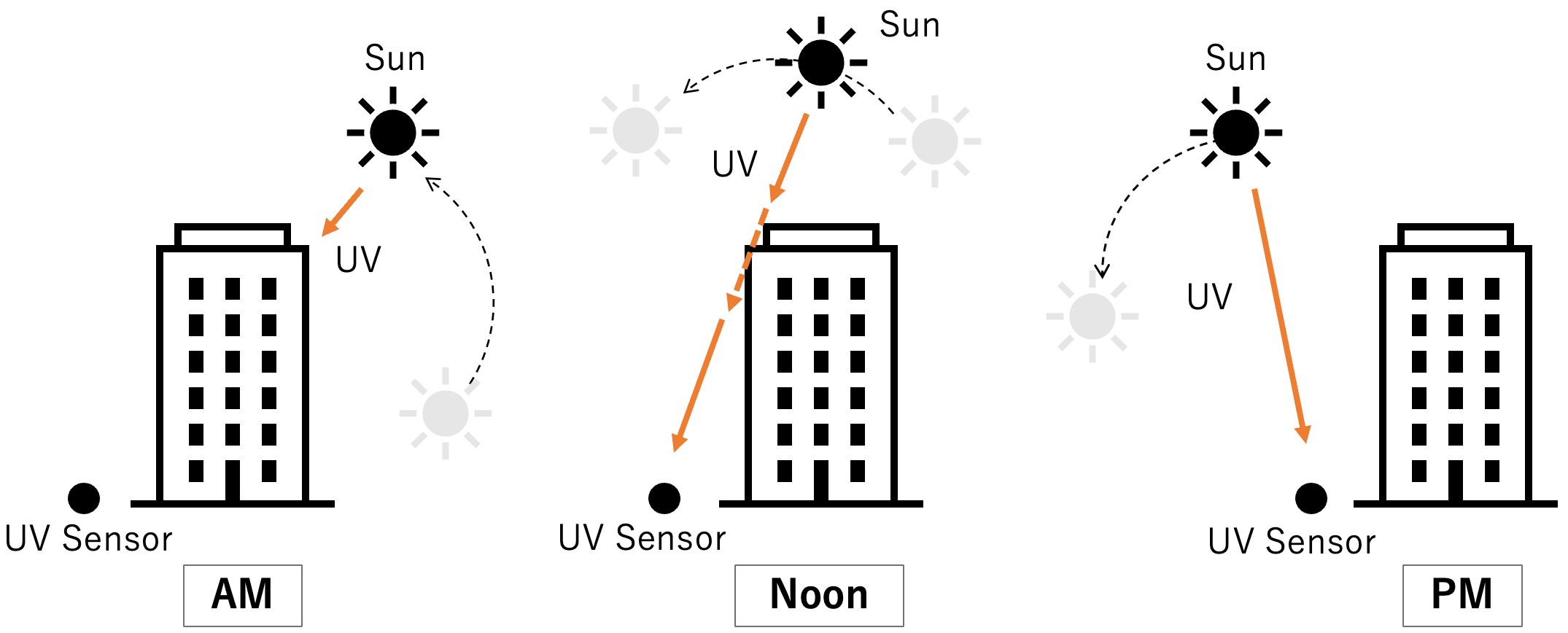}
    \caption{The assumption of a shady (at AM) and sunny (at PM) place.}
    \label{fig:figure47}
\end{figure}

\begin{itemize}
    \item Date: Daytime on October 18 in 2021
    \item location: Two locations in sun to shade and two locations in shade to sun as shown in Figure \ref{fig:figure47}.
    \item description: UV Index and GNSS-signal-related values were measured using the device shown in Figure \ref{fig:figure42}.
\end{itemize}

\begin{figure}
    \centering
    \includegraphics[width=\linewidth]{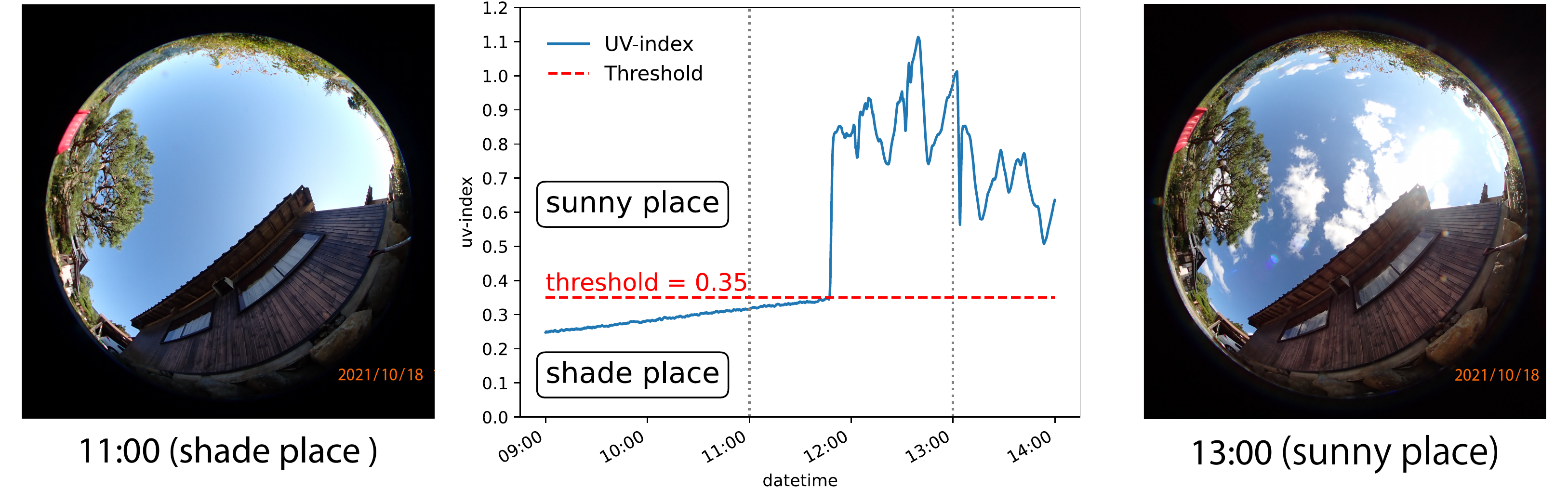}
    \caption{UV Index on October 18. This graph shows that the UV Index did not reach 0 even in the shade. On the basis of values in the graph, we defined shade as a UV Index<0.35.}
    \label{fig:figure413}
\end{figure}

The data format from one of the UV sensors was corrupted when the sensor moved from sun to shade, so it was excluded. Therefore, one sun-to-shade data and two shade-to-sun data were combined. The UV Index is shown in Figure \ref{fig:figure413}. The index values for the sun and shade were slightly larger than 0, unlike the indoor values. Based on the above data, we set a UV Index of 0.35 as the threshold value. In other words, when the UV Index is smaller than 0.35, the area is considered to be shaded.

\section{GNSS-signal-based Detection Method}
\label{sec:approach}

\begin{figure}[tbh]
 \centering
 \includegraphics[width=0.9\linewidth]{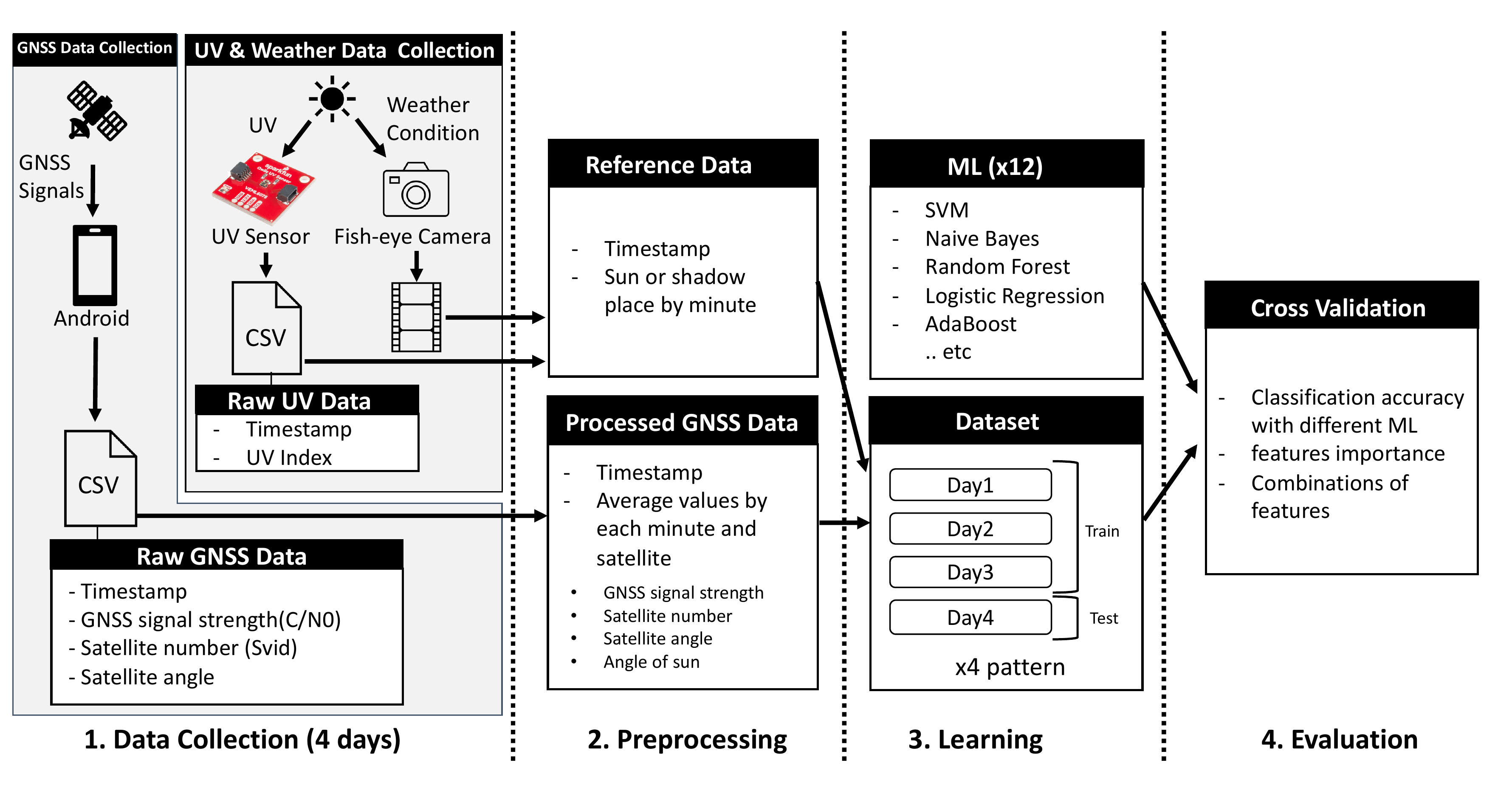}
 \caption{Overview of Detection Method.}
 \label{fig:figure40}
\end{figure}

Fig. \ref{fig:figure40} is an overview of the detection method. The purpose of the estimation was to verify whether it is possible to estimate shade and sun from the GNSS signal data in as simple an environment as possible. In addition to the GNSS data, sun angle was used as data in this experiment. Since machine learning methods are not the main topic of this study, we used multiple methods and used the overall results to evaluate the accuracy of detection. To prevent the results from becoming biased by biased data, we prepared four types of training and test data and summarized the results for each.

\subsection{Measurement Tools}
\begin{figure}[tbp]
    \centering
    \begin{minipage}[b]{0.55\linewidth}
        \centering
        \includegraphics[width=\linewidth]{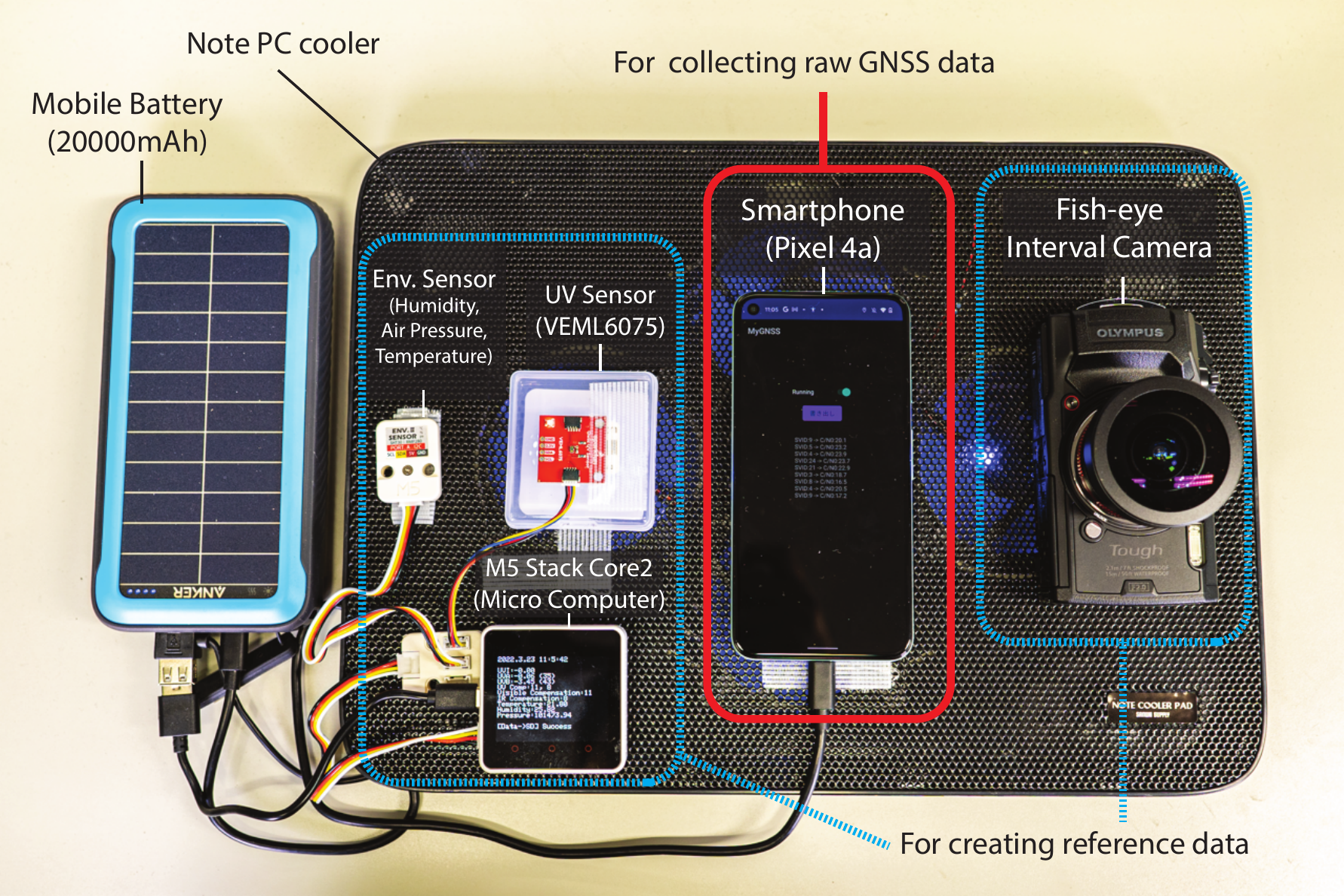}
        \caption{Setup of measurement tools.}
        \label{fig:figure42}
    \end{minipage}
    \hfill
    \begin{minipage}[b]{0.44\linewidth}
        \centering
        \includegraphics[width=\linewidth]{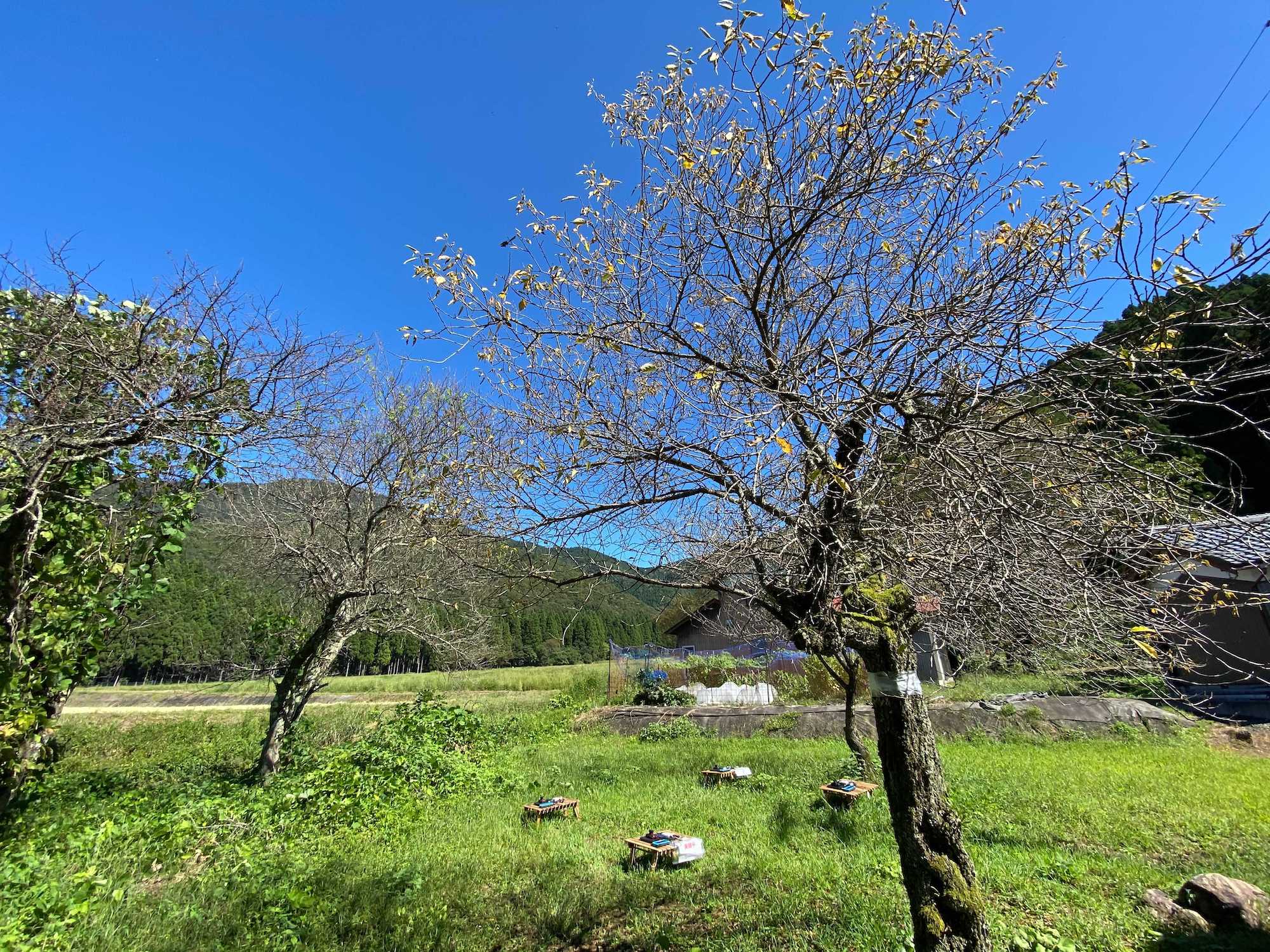}
        \caption{Experimental location in a rural area.}
        \label{fig:figure41}
    \end{minipage}
\end{figure}

The setup of the measurement tools is shown in Fig.~\ref{fig:figure42}. These tools were mainly used for collecting raw GNSS data and ground-truth data.

For collecting raw GNSS signals during a day, we created a logger application based on GnssLogger, which is open-source software released by Google~\footnote{\url{https://developer.android.com/guide/topics/sensors/gnss}}. The application collects National Marine Electronics Association (NMEA) 0183 format data, as raw GNSS signals (see Section~\ref{sec:raw_gnss_data}), via \verb|OnNmeaMessageListener| on Android. The collected raw GNSS data was saved in a text file. Moreover, we used Google Pixel serise smartphone (Android OS version 9.0 or higher) as GNSS data collection platform.

For collecting ground-truth data on sunny and shade places, we collected UV data and 180-degree images of the upper air. The UV data were collected by VEML6075. The UV sensor module could collect UV-A, B, and Index. M5Stack Core2, which is an ESP32-based microcomputer, received the collected UV values every second from the sensor module via the I2C communication protocol and saved them as a CSV file on a micro SD card in the M5Stack Core2. M5Stack Core2 has a real-time clock (RTC) module; thus, the collected sensor data can be saved together with the timestamp. Moreover, the RTC module is initialized before data collection by using the network time protocol (NTP). As a 180-degree camera, we used a compact digital camera (Tough TG-6 from Olympus Corporation) with a fisheye converter lens. During the data collection, the camera took upper air photos every 1 minute by using its interval shooting function.

From a preliminary experiment, we faced the battery capacity issue with the M5 Stack and the overheating issue of the smartphone and M5Stack when they were under the hot sun. To solve this problem, we used a mobile battery (PowerCore Solar 2000, Anker) and air cooling fans (USB fan for a notebook pc). Moreover, we confirmed through preliminary experiments that these devices did not affect the experimental results. 

\subsection{Dataset}
\label{sec:dataset}

The data were collected in a large area in the town of \verb|${PREF_NAME}|, \verb|${COUNTRY_NAME}|, to avoid the effect of multipath in urban areas. A photo of the experimental site is shown in Fig. \ref{fig:figure41}. The experiment was conducted for about 12 hours from sunrise to sunset.

We conducted the simplest experiment under clear skies. As a control experiment, we also experimented with the shade indoors and compared the results.

\begin{itemize}
\item Date: daytime of September 21 (Day-1), 25 (Day-2), 27 (Day-3), and 30 (Day-4) in 2021
\item Location: four outdoor and one indoor location.
\item description: UV Index and GNSS-signal-related values were measured using the device shown in Figure\ref{fig:figure42}.
\end{itemize}

\subsection{Data Extraction for Machine Learning}
The preprocessed dataset for machine learning contained ground-truth data and extracted features of GNSS data. Section~\ref{sec:feature_extraction} describes the feature extraction method, and Section~\ref{sec:ground_truth} describes the method of creating ground-truth labels from the UV values.

\subsubsection{Feature Data}
\label{sec:feature_extraction}
We extracted 15 features from these raw GNSS signals, as shown in Table~\ref{tab:Feature_Data}: average C/N0 ($S(t)$), azimuth and elevation of each satellite ($A_{SAT}(t)$ and $E_{SAT}(t)$), azimuth and elevation of the sun ($A_{SUN}(t)$ and $E_{SUN}(t)$), and values of $A_{SAT}(t)$ and $E_{SAT}(t)$, $A_{SUN}(t)$, $E_{SUN}(t)$ for one minute before and after the time, i.e. values in $t-1$ and $t+1$. The values of the GNSS signals from each satellite were averaged per minute. Each satellite was identified by its SVID. The solar position was calculated using the \verb|get_solarposition| function in \verb|pvlib| (Version 0.9.1)~\cite{Holmgren2018}. \verb|pvlib| is a python library that calculates the angle of the sun from the latitude, longitude, and time of the location. We treated the features listed in Table~\ref{tab:Feature_Data}.

\begin{table}[tb]
    \centering
    \caption{Features for machine learning.}
    \begin{tabular}{l l l} 
    \toprule
        Feature set & Feature name & Description \\
    \midrule 
        CN0 & $S(t)$ & C/N0 value (GNSS strength) at time $t$\\
    \addlinespace[2mm] 
        SAT & $A_{SAT}(t)$ & Azimuth value of a satellite at time $t$\\
            & $E_{SAT}(t)$ & Elevation value of a satellite at time $t$\\
    \addlinespace[2mm] 
        SUN & $A_{SUN}(t)$ & Azimuth value of the sun at time $t$ \\
            & $E_{SUN}(t)$ & Elevation value of the sun at time $t$ \\
    \addlinespace[2mm] 
        $\Delta t$& $S(t-1)$ & C/N0 value (GNSS strength) at time $t-1$\\
            & $S(t+1)$ & C/N0 value (GNSS strength) at time $t+1$\\
            & $A_{SAT}(t-1)$ & Azimuth value of a satellite at time $t-1$\\
            & $A_{SAT}(t+1)$ & Azimuth value of a satellite at time $t+1$\\
            & $E_{SAT}(t-1)$ & Elevation value of a satellite at time $t-1$\\
            & $E_{SAT}(t+1)$ & Elevation value of a satellite at time $t+1$\\
            & $A_{SUN}(t-1)$ & Azimuth value of the sun at time $t-1$ \\
            & $A_{SUN}(t+1)$ & Azimuth value of the sun at time $t+1$ \\
            & $E_{SUN}(t-1)$ & Elevation value of the sun at time $t-1$ \\
            & $E_{SUN}(t+1)$ & Elevation value of the sun at time $t+1$ \\
    \bottomrule
    \end{tabular}
    \label{tab:Feature_Data}
\end{table}

To take advantage of the continuity of the data, we also included $t+1$ and $t-1$ data for time $t$ as features. Therefore, the number of features for t was 15. The experiment was conducted over a half-day period in which the sun was out on four days, and the number of data sets including per-minute averages exceeded 37,000.

\subsubsection{Ground-truth Data}
\label{sec:ground_truth}

A UV sensor (VEML6075) was used to measure the ground truth. Although it provides a variety of information (e.g., temperature, humidity, and air pressure), we used only the UVI information, which is generally used for determining the intensity of UV light. The sun-shade threshold values shown in Section \ref{subsec:def_sun} were used to determine whether the measurement point was in the sun or in the shade. The result was used as the ground-truth score.


%
%
%
%
%
%
%
%

\section{Evaluation and Results}
\label{sec:evaluation}
The goal of this study is to passively detect sunny and shady places by using GNSS signals that can be received by an off-the-shelf mobile device like a smartphone. To achieve it, we attempted to answer three research questions: 
\begin{itemize}
\item RQ1: How accurately can shady and sunny places be detected from GNSS signals picked up by a smartphone?
\item RQ2: Which features are effective for detecting shady and sunny places? 
\item RQ3: Which combination of features is most effective?
\end{itemize}

Below, section~\ref{sec:evaluation_procedure} describes the procedure of the evaluation, including the machine learning methods used and the datasets on sunny-and-shady places. Section~\ref{sec:rq1} describes the performance evaluation of the place classification task that used multiple machine learning methods. Section~\ref{sec:rq2}investigates the feature importance. Section ~\ref{sec:rq3} examines the effect of combining features on the classification performance.

\subsection{Evaluation Procedure}
\label{sec:evaluation_procedure}
Our hypothesis is that the GNSS signal and sunlight have similar characteristics, so it is possible to use the GNSS signal to estimate the sunshine state (i.e., sunny or shady) at a location. To examine this hypothesis, we took a supervised machine learning approach. In particular, we created and evaluated a machine learning model for estimating the sunshine state from GNSS signals received by a smartphone.

Section~\ref{sec:evaluation_procedure_dataset} describes dataset on the sunshine state and GNSS data. Section~\ref{sec:evaluation_procedure_ml} describes the 12 machine learning methods of this evaluation.

\subsubsection{Training and test data}
\label{sec:evaluation_procedure_dataset}
We used the dataset of the sunshine state and GNSS signals mentioned in Section 1. The dataset includes four days’ worth of data. In order to conduct 4-fold cross-validations, it is split into four dataset patterns by date, as shown in Table~\ref{tab:dataset-pattern}.

The training and test datasets contain three- and one-day data, respectively. Because the UV-Index values can be inferred from the data before and after the recorded time, the training and test data should be on different days. In addition, there is a possibility that the results may be biased depending on the choice of schedule. Therefore, we performed our evaluations with four dataset patterns. In addition, we equalized the amount of training data for sunny and shade places to eliminate any bias. 

\begin{table}[tbh]
    \centering
    \caption{Training and test dataset.}
    \begin{tabular}{l l l}
    \toprule
        Pattern & Training Data & Test Data \\
    \midrule
        Pattern-1 & Day-1, 2, and 3 (n=27,900) & Day-4 (n=9,725) \\
        Pattern-2 & Day-1, 2, and 4 (n=27,364) & Day-3 (n=9,627) \\
        Pattern-3 & Day-1, 3, and 4 (n=28,206) & Day-2 (n=8,738) \\
        Pattern-4 & Day-2, 3, and 4 (n=28,060) & Day-1 (n=9,814) \\
    \bottomrule
    \end{tabular}
    \label{tab:dataset-pattern}
\end{table}


\subsubsection{Machine Learning Method}
\label{sec:evaluation_procedure_ml}

The task was a binary classification of sunny or shady places from GNSS signals. We used twelve machine learning methods that can conduct binary classification, i.e., Support Vector Machine (SVM) with four kernels (Linear~\cite{Linear}, Polynomial~\cite{polynomial}, RBF~\cite{RBF}, and Sigmoid~\cite{sigmold}), Decision Tree~\cite{DT}, Random Forest~\cite{random}, Logistic Regression~\cite{logistic}, AdaBoost(Adaptive Boosting)~\cite{ada}, Naive Bayes~\cite{bayes}, K-Neighbors Classifier~\cite{K}, Quadratic Discriminant Analysis (QDA)~\cite{qda}, and Linear Discriminant Analysis (LDA)~\cite{lda}.

Note that our aim was not to select the most suitable algorithm for optimization, but rather to investigate the framework itself. The purpose of performing the regression with twelve different methods was not to select the best method for the dataset, but to show the performance of the various learning methods in general.
There are a wide variety of binary classification methods, and which study will be more efficient depends on the characteristics of each.
there have been many studies in the past on the differences in results from each learning method~\cite{comparison,comparison2}.
In this study, we compared the results of 12 learning methods implemented in Scikit-learn 0.24.1 for Python 3.8.8, and discussed the overall estimation accuracy independent of the learning method.

\subsection{Accuracy of detection}
\label{sec:rq1}
To address RQ1, we performed a 4-fold cross-validation using the dataset (including four patterns) on twelve machine learning models. Table~\ref{fig:figure51} shows the results. Each accuracy, recall, precision, and f1-score is the average of four patterns (pattern-1 to -4 in section~\ref{sec:evaluation_procedure_dataset}).

Eleven of the twelve methods had classification accuracies of more than 90\%. This result indicated that the sunshine estimation on the dataset was successful. In particular, SVM (RBF) performed the best (f1: 0.972, recall: 0.972, precision: 0.972). On the other hand, SVM (Sigmoid) performed the worst (accuracy: 0.878, f1: 0.878, recall: 0.878, precision: 0.882). The nonlinear regression methods had better predictive performance than the linear one; thus, the features could be used to make shade and sun decisions under complex conditions.

\begin{table}[bt]
    \centering
    \begin{minipage}{0.7\linewidth}
        \centering
        \caption{Result of 4-fold cross-validation.}
        \begin{tabular}{l r r r r } 
        \toprule
        Method & Accuracy & Recall & Precision & f1-score \\ 
        \midrule
        SVM (RBF)       & 0.972& 0.972& 0.972 & 0.972 \\ 
        QDA             & 0.972& 0.973& 0.972 & 0.972\\ 
        Naive Bayes     & 0.962& 0.962& 0.962 & 0.962\\ 
        SVM (Polynomial)& 0.959& 0.959& 0.960 & 0.959 \\ 
        Decision Tree & 0.955 & 0.955& 0.956 & 0.955\\ 
        K-nearest Neighbor & 0.950& 0.952& 0.949 & 0.950\\
        Random Forest   & 0.948& 0.948& 0.949 & 0.948 \\ 
        Logistic Regression & 0.937& 0.939& 0.940 & 0.937 \\ 
        SVM (Linear)    & 0.935& 0.937& 0.938 & 0.935\\ 
        LDA             & 0.932& 0.936& 0.935 & 0.932\\
        AdaBoost        & 0.904& 0.906& 0.907 & 0.904\\ 
        SVM (Sigmoid)   & 0.878& 0.880& 0.881 & 0.878\\ 
        \bottomrule
        \end{tabular}
        \label{fig:figure51}
    \end{minipage}
    \hfill
    \begin{minipage}{0.25\linewidth}
        \centering
        \caption{Top-five most important features.}
        \begin{tabular}{l r} 
        \toprule
        Feature name & Weight  \\ 
        \midrule
        $S(t)$ & $0.394 \pm{0.007}$ \\ 
        $E_{SUN}(t)$ & $0.072 \pm{0.002}$\\ 
        $S(t+1)$ & $0.032 \pm{0.004}$ \\ 
        $A_{SUN}(t)$ & $0.029 \pm{0.002}$ \\ 
        $S(t-1)$ & $0.026 \pm{0.004}$ \\ 
        \bottomrule
        \end{tabular}
        \label{fig:figure52}
    \end{minipage}
\end{table}

\subsection{Feature importance of detection model}
\label{sec:rq2}
On the basis of the results described in Section~\ref{sec:rq1}, we investigated the feature importance of the machine learning model. Particularly, we chose the RBF SVM-based classification model as the target of the investigation. Table \ref{fig:figure52} shows the results of the top-five most important features of the classification model. The feature importance was obtained using Eli5~\footnote{\url{https://eli5.readthedocs.io/en/latest/overview.html}}. As shown in the table, the features affect the detection accuracy in the order $S(t)$, $S(t+1)$, $S(t-1)$, $E_{SAT}(t)$, and $E_{SUN}(t)$. $S(t)$, $S(t+1)$, and $S(t-1)$ are the strengths of the signal from a satellite. This result indicates that the signal strength is the most important feature for detecting sunny and shady places. $E_{SAT}(t)$ and $E_{SUN}$ slightly contributed the performance, each with weight 0.01. The next section~\ref{sec:rq3} investigates how the features and their combination contribute to the classification performance.

\subsection{Feature combinations for improving performance}
\label{sec:rq3}
To investigate the effect of feature combinations on the accuracy of machine learning, we used four important features of the SVM (RBF)-based classification model: C/N0, angle of satellites, angle of the sun, and features values before and after the time of the feature values. We used SVM (RBF) and compared 14 combinations of features in terms of accuracy, recall, precision, and f1-score.

\begin{figure}[tb]
    \centering
    \includegraphics[width=\linewidth]{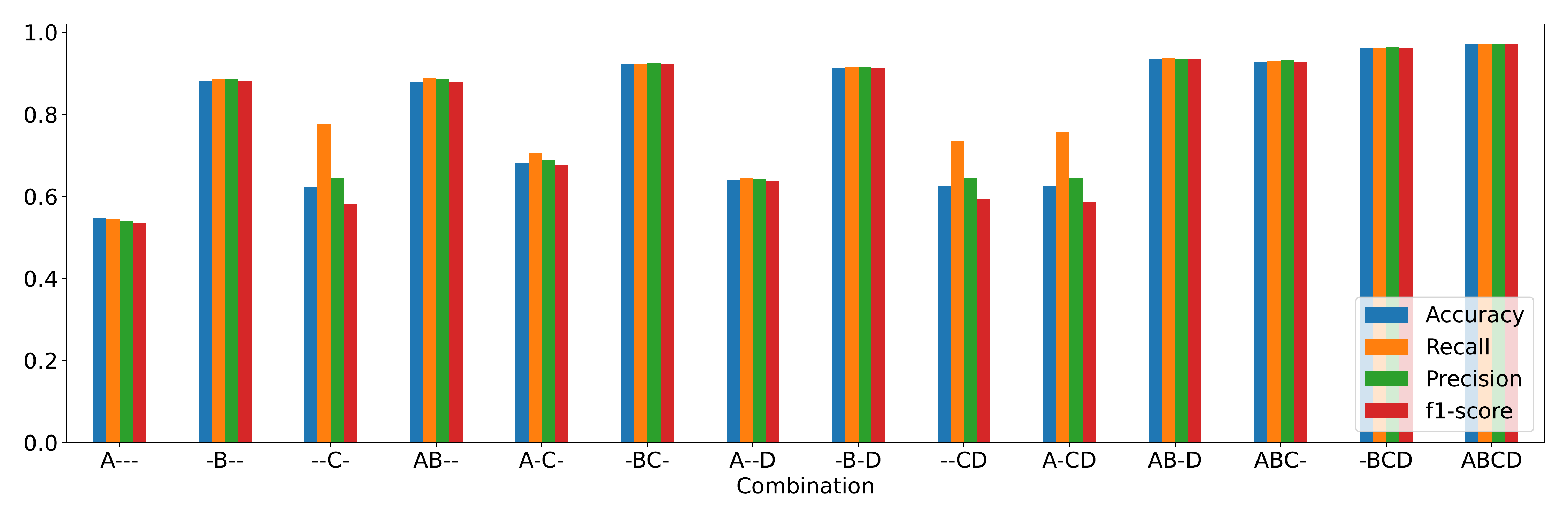}
    \caption{Detection accuracy for each combination. {\bf A}, {\bf B}, {\bf C}, {\bf D}, and {\bf -} mean angle of satellites (SAT features), C/N0 (CN0 feature), angle of the sun (SUN features), data feature values before and after the target time ($\Delta t$ features), and not applicable, respectively.}
    \label{fig:figure521}
\end{figure}

The results are shown in Fig.~\ref{fig:figure521}. \verb|A|, \verb|B|, \verb|C|, \verb|D|, and \verb|-| on the x-axis mean the angle of satellites (SAT features), C/N0 (CN0 feature), angle of the sun (SUN features), features value before and after the time of the feature values ($\Delta t$ features), and not applicable (NA), respectively. For example, \verb|A---| means the classification model used the angle of satellites and \verb|ABCD| means that the classification model used all features.

All predictions that included C/N0 had accuracies above 0.8. On the other hand, all predictions that did not include C/N0 did not exceed 0.7, indicating that C/N0 is important. Moreover, the predictions that included all factors were the most effective, indicating that each factor was effective in the prediction, but that satellite angle data were less influential than the other data.

\verb|A| (satellite angle) alone seems to be near 0.5, however, this task is a binary classification task, then it has no predictive ability because 0.5 is the same performance as a random decision. 
On the other hand, \verb|B| and \verb|C| differ from a random decision (0.5) and have a slight signal by themselves.
The accuracy of the case of all feature sets (\verb|ABCD|) was 0.972, therefore other feature sets also slightly contribute to improving prediction accuracy.
Moreover, the recall is particularly high when \verb|C| is included, which suggests that the inclusion of \verb|SUN| may increase the tolerance to noise and unobserved data conditions that cannot be determined by GNSS data alone.


\section{Discussion}
\label{sec:discussion}
\subsection{Prediction in Other Conditions}
We tested the applicability of our method to other locations and on different days. In the following month (November 4th, 2021), we conducted another experiment at a different location, approximately 200 meters away from the experimental validation reported in Section 5. We applied the predictive model trained on the dataset presented in Section~\ref{sec:evaluation} to the measured data to determine whether the site was in the shade or in the sun. All features (\verb|A|, \verb|B|, \verb|C|, and \verb|D|) were used, and the best-performing SVM (RBF) was used as the training method.

The results showed high {\bf \{accuracy, precision, recall, and f1-score\} =\{0.91, 0.92, 0.92, 0.91\}}. Although the values are lower than those in Section \ref{sec:evaluation}, they were sufficiently high even though the climate conditions and the measured values were different.

\subsection{Increasing Accuracy}
The weather forecast provides information such as estimated temperature, humidity, amount of cloud, UV Index, and air pressure, around the place. By creating a classification model combining with GNSS signal and these weather forecast data, potentially, the detection accuracy and robustness of the model will be increased.
In this paper, we focused on the average value of the GNSS signals collected in one minute; we can expect a further increase in accuracy by performing machine learning using the difference in the strength of GNSS signals received in one minute before and after the average value.

\subsection{Potential Applications}
The following might be a practical application of our estimation method. By switching tabs, measurement and recording screens can be alternately displayed. The time spent in the sun per day could be shown as a bar graph on the recording screen. The purpose would be to encourage people to voluntarily go out in the sun by showing on the same graph the optimal amount of UV exposure per day for their health when it becomes possible to estimate the amount of UV exposure. Considering that the amount of sunshine and the UV Index can be estimated in real-time on a smartphone, other information can also be used to improve accuracy. In addition, we used 1-minute average data in our study, meaning that our method is not suitable for estimating sunshine while the user is moving. In the future, we should investigate the effect of averaging over a shorter period on the estimation.

\subsection{Limitations}
In this paper, we evaluated the performance of the binary classification task (i.e., sunny or shady place) at the same place and season by using our proposed method. As shown in Section~\ref{sec:evaluation}, the classification accuracy has been more than 90\% accuracy in the condition above. 
However, our evaluation is conducted in the same place and season, it is not known whether our proposed method is applicable in the other place and season. To address the limitation, we need to collect data in multiple places and evaluated the classification performance. 

Moreover, our method needs 1 minute to detect the sunny or shady conditions because of the features that depend on the latest one minute. To shorten the duration until the detection, we need to optimize the feature selection. 
Related to the above, another limitation is the performance of the smartphone is carried by a user. As a benchmark, we have collected raw GNSS data without noise as possible in this paper. Therefore, the GNSS data is collected in a fixed place and same season.  However, in a real use case, a user tends to be carrying a smartphone in the pocket or bag. To evaluate the applicability of our method, we need to conduct our study in a real-world setting. Performance evaluations in these noisy conditions are our future work.

\section{Conclusion}
\label{sec:conclusion}

Because both excessive and insufficient UV exposure leads to various health problems, detecting the spent time in sunny and shady place sunny and shady places is important to prevent disease.
In this paper, we proposed a method that allows us to detect sunny and shadow places using a GNSS strength that can be collected on off-the-shelf smartphones.  
Our method is based on the characteristic that both GNSS signal and UV index are attenuation by screening objects. 
We created a machine learning model to classify sunny or shady places with GNSS-related signals (i.e., angle and signal strength of each satellite, and the sun angle calculated from the location) that are collected from an off-the-shelf smartphone.
As a dataset of sunny and shady places, we collected show/sunny places labeled data containing GNSS-related raw data for four days and five places. Based on the dataset, we created classification models with twelve machine learning methods and different feature combinations.
Our performance evaluation showed that the model can classify shadow and sunny places by a minute with more than 95\% accuracy. Moreover, feature importance analytics indicated the signal strength of C/N0 at the moment is the most important feature. In the analytics of feature combination, we found that our method could detect sunny and shady places with more than 80\% accuracy even when using only C/N0 features. Adding the angle of the sun and delta feature (one minute before and after of features at the target time) increased nearly 10\% of accuracy.

\bibliographystyle{ACM-Reference-Format}
\bibliography{citations}





\end{document}